\documentclass[useAMS,usenatbib]{mn2e}
\usepackage{graphicx,natbib,times,amssymb}
\usepackage[T1]{fontenc} 
 \usepackage{aecompl} 
\title[AGN flickering]{Active galactic nuclei flicker: an observational estimate of the duration of black hole growth phases of $\sim10^{5}$ years}
\author[Kevin Schawinski et al.]{
 \parbox[t]{18cm}{
 Kevin Schawinski$^{1}$\thanks{E-mail: kevin.schawinski@phys.ethz.ch, Twitter: @kevinschawinski}, Michael Koss$^{1}$, Simon Berney$^{1}$ and Lia F. Sartori$^{1}$\\
 }\\
$^{1}$Institute for Astronomy, Department of Physics, ETH Zurich, Wolfgang-Pauli-Strasse 27, CH-8093 Zurich, Switzerland\\
}

\begin{document}

\newcommand\aj{{AJ}}%
\newcommand\actaa{{Acta Astron.}}%
\newcommand\araa{{ARA\&A}}%
\newcommand\apj{{ApJ}}%
\newcommand\apjl{{ApJ}}%
\newcommand\apjs{{ApJS}}%
\newcommand\ao{{Appl.~Opt.}}%
\newcommand\apss{{Ap\&SS}}%
\newcommand\aap{{A\&A}}%
\newcommand\aapr{{A\&A~Rev.}}%
\newcommand\aaps{{A\&AS}}%
\newcommand\azh{{AZh}}%
\newcommand\baas{{BAAS}}%
\newcommand\caa{{Chinese Astron. Astrophys.}}%
\newcommand\cjaa{{Chinese J. Astron. Astrophys.}}%
\newcommand\icarus{{Icarus}}%
\newcommand\jcap{{J. Cosmology Astropart. Phys.}}%
\newcommand\jrasc{{JRASC}}%
\newcommand\memras{{MmRAS}}%
\newcommand\mnras{{MNRAS}}%
\newcommand\na{{New A}}%
\newcommand\nar{{New A Rev.}}%
\newcommand\pra{{Phys.~Rev.~A}}%
\newcommand\prb{{Phys.~Rev.~B}}%
\newcommand\prc{{Phys.~Rev.~C}}%
\newcommand\prd{{Phys.~Rev.~D}}%
\newcommand\pre{{Phys.~Rev.~E}}%
\newcommand\prl{{Phys.~Rev.~Lett.}}%
\newcommand\pasa{{PASA}}%
\newcommand\pasp{{PASP}}%
\newcommand\pasj{{PASJ}}%
\newcommand\qjras{{QJRAS}}%
\newcommand\rmxaa{{Rev. Mexicana Astron. Astrofis.}}%
\newcommand\skytel{{S\&T}}%
\newcommand\solphys{{Sol.~Phys.}}%
\newcommand\sovast{{Soviet~Ast.}}%
\newcommand\ssr{{Space~Sci.~Rev.}}%
\newcommand\zap{{ZAp}}%
\newcommand\nat{{Nature}}%
\newcommand\iaucirc{{IAU~Circ.}}%
\newcommand\aplett{{Astrophys.~Lett.}}%
\newcommand\apspr{{Astrophys.~Space~Phys.~Res.}}%
\newcommand\bain{{Bull.~Astron.~Inst.~Netherlands}}%
\newcommand\fcp{{Fund.~Cosmic~Phys.}}%
\newcommand\gca{{Geochim.~Cosmochim.~Acta}}%
\newcommand\grl{{Geophys.~Res.~Lett.}}%
\newcommand\jcp{{J.~Chem.~Phys.}}%
\newcommand\jgr{{J.~Geophys.~Res.}}%
\newcommand\jqsrt{{J.~Quant.~Spec.~Radiat.~Transf.}}%
\newcommand\memsai{{Mem.~Soc.~Astron.~Italiana}}%
\newcommand\nphysa{{Nucl.~Phys.~A}}%
\newcommand\physrep{{Phys.~Rep.}}%
\newcommand\physscr{{Phys.~Scr}}%
\newcommand\planss{{Planet.~Space~Sci.}}%
\newcommand\procspie{{Proc.~SPIE}}%
\newcommand\helvet{{Helvetica~Phys.~Acta}}%

\def\Chandra{\textit{Chandra}}
\def\XMM{\textit{XMM-Newton}}
\def\Swift{\textit{Swift}}

\def\OI{[\mbox{O\,{\sc i}}]~$\lambda 6300$}
\def\OIII{[\mbox{O\,{\sc iii}}]~$\lambda 5007$}
\def\SII{[\mbox{S\,{\sc ii}}]~$\lambda \lambda 6717,6731$}
\def\NII{[\mbox{N\,{\sc ii}}]~$\lambda 6584$}

\def\Ha{{H$\alpha$}}
\def\Hb{{H$\beta$}}

\def\NIIHa{[\mbox{N\,{\sc ii}}]/H$\alpha$}
\def\SIIHa{[\mbox{S\,{\sc ii}}]/H$\alpha$}
\def\OIHa{[\mbox{O\,{\sc i}}]/H$\alpha$}
\def\OIIIHb{[\mbox{O\,{\sc iii}}]/H$\beta$}

\def\Ebmv{E($B-V$)}
\def\LOIII{$L[\mbox{O\,{\sc iii}}]$}
\def\Ledd{${L/L_{\rm Edd}}$}
\def\LOIIIs4{$L[\mbox{O\,{\sc iii}}]$/$\sigma^4$}
\def\LOIIIMbh{$L[\mbox{O\,{\sc iii}}]$/$M_{\rm BH}$}
\def\Mbh{$M_{\rm BH}$}
\def\Msigma{$M_{\rm BH} --- \sigma$}
\def\Ms{$M_{\rm *}$}
\def\Msun{$M_{\odot}$}
\def\Msunyr{$M_{\odot}yr^{-1}$}

\def\ergs{$~\rm erg~s^{-1}$}
\def\kms{$~\rm km~s^{-1}$}

\def\galfit{\texttt{GALFIT}}
\def\multidrizzle{\texttt{multidrizzle}}

\def\sersic{S\'{e}rsic}

\date{}

\pagerange{\pageref{firstpage}--\pageref{lastpage}} \pubyear{2015}

\maketitle

\label{firstpage}

\begin{abstract}
We present an observational constraint for the typical active galactic nucleus (AGN) phase lifetime. The argument is based on the time lag between an AGN central engine switching on and becoming visible in X-rays, and the time the AGN then requires to photoionize a large fraction of the host galaxy. Based on the typical light travel time across massive galaxies, and the observed fraction of X-ray selected AGN without AGN-photoionized narrow lines, we estimate that the AGN phase typically lasts $\sim10^{5}$ years. This lifetime is short compared to the total growth time of $10^{7}-10^{9}$ years estimated from e.g. the Soltan argument and implies that black holes grow via many such short bursts and that AGN therefore "flicker" on and off. We discuss some consequences of this flickering behavior for AGN feedback and the analogy of X-ray binaries and AGN lifecycles.
\end{abstract}


\section{Introduction}
\label{sec:intro}
When the massive black holes in galaxy centers accrete, they become visible as quasars or active galactic nuclei (AGN). The energy liberated by accretion episodes is thought to be a critical regulatory mechanism in galaxy evolution \citep[e.g.][]{1988ApJ...325...74S, 1998A&A...331L...1S}. But how exactly does this connection operate? Amongst the most basic unknowns is the timescale on which the AGN phase operates. Indirect measurements of the typical AGN phase are on the order of $10^{7} - 10^{9}$ years \citep[e.g.][]{2001ApJ...547...12M, 2004MNRAS.351..169M}. Taking the local relic black hole mass density and comparing it to the total light emitted by quasars and assuming a reasonable radiative efficiency yields such long ($10^{7} - 10^{9}$ year) \textit{total} accretion timescales \cite{1982MNRAS.200..115S, 2002MNRAS.335..965Y, 2004MNRAS.351..169M}. This does not constrain however whether the entire mass growth consists of a single accretion phase, or is broken up into shorter phases.

AGN variability has been observed on short time scales all the way out to multiple decades \citep[e.g.][]{1997ARA&A..35..445U}. The \textit{natural time scales} of the AGN central engine however are significantly longer than human time scales. This leaves a major gap on the timescale domain on which we can easily study AGN variability: we can see AGN vary on human timescale, and we can constrain the total growth time using statistical arguments.

In this paper, we present an \textit{observational argument} for why nearby AGN -- and perhaps all AGN --  switch on rapidly ($\sim 10^{4}$ years) and "stay on'" for $\sim 10^{5}$ years before switching off. In order to reconcile this result with previous measurements of AGN lifetimes, we argue that AGN \textit{flicker} in a $\sim 10^{5}$ year cycle resulting in the total  $10^{7-9}$ year lifetime. We therefore begin to fill in AGN variability between the two extremes.
 
Breaking up the accretion phases of AGN into many short $\sim 10^{5}$ year phases has implications beyond the fact that massive black holes assemble their mass via a large number of individual bursts. Many short phases also means that the AGN central engine swings back and forth between different accretion states. During high-Eddington phases, the AGN will produce predominantly radiative energy and will be visible as a classical quasar or AGN. During the low-Eddington phases, the bulk of the energy is in the form of kinetic energy \citep{2007A&ARv..15....1D, 2012NewAR..56...93A}. In this low-Eddington phase, the AGN is harder to detect, but may still inject (kinetic) energy into the host galaxy. Thus, a full accounting of the AGN lifecycle and the different modes in which an AGN interacts the host galaxy may be vital to a full description of AGN feedback and the galaxy-black hole connection.  Ultimately, we will need constraints on the full power spectrum of AGN from days to several Gyrs to understand how black holes influence their surrounding galaxies.

In this paper, we outline the argument for a short AGN lifetime and discuss some implications for our understanding of the galaxy-black hole connection and feedback.


\section{Analysis}
\label{sec:analysis}
In this Section, we outline our new framework of how existing observations can constrain the duration of an individual AGN phase to $\sim10^{5}$ years, present these observations, and then give an order of magnitude calculation.

\subsection{Framework}
Consider the case of a central black hole in a star-forming galaxy. When the black hole begins accreting, it become visible virtually instantaneously as a nuclear source. The accretion disk and surrounding material begin emitting at optical and ultraviolet wavelengths, and the hard X-rays will escape the nuclear region even at very high levels of obscuration. As the AGN phase continues in time, photons from the central engine begin to photoionize the interstellar medium (ISM) of the host galaxy. Due to the geometry of obscuring medium, or torus, the photoionized region usually manifests itself as an ionization cone \citep{1998ApJS..117...25M}. Once these galaxy-scale photoionized \textit{narrow line regions} are established, the host galaxy can be identified as an AGN as the central engine dominates the photoionization budget of the host galaxy \citep{1981PASP...93....5B, 1987ApJS...63..295V}. \textbf{The narrow line region (NLR) is therefore the part of the host galaxy interstellar medium whose emission lines are driven by AGN photoionization.} The most prominent AGN line seen in optical spectra is the \OIII\ line, though usually a series of line ratio diagnostics is used. Higher ionization coronal lines are often also found.

From this setup, it follows that there is a significant \textit{time delay} between the start of accretion in the AGN central engine, and the time the host galaxy exhibits AGN-like line ratios. To zeroth order, this time delay is on the light travel time from the nucleus to the half-light radius of the host galaxy. Naturally, the properties of the ISM would be an important first order effect in this time delay as the complex physics of photoionization kicks in, but we believe that estimating the switch-on time using the light travel time across the host galaxy is the dominant effect. As the AGN photoionization travels across the host galaxy, AGN lines become stronger and more apparent over those driven by star formation and shocks in the galaxy, and to the host galaxy spectrum may begin to move into the "composite" region on emission line diagrams. Ultimately the AGN will photoionize the ISM and the host galaxy will be classified as a regular AGN. 

The non-trivial light travel time of the AGN radiation across the host galaxy thus constrains the \textit{lifetime of the AGN phase} in a statistical sense.
Assuming that those AGN host galaxies in a (hard) X-ray selected AGN sample whose spectra show either pure star-formation or composite AGN+star formation lines are those which just switched on, then the combination of the fraction and the light travel time across the host galaxy yields a lifetime for the AGN phase:
\begin{equation}
t_{\rm AGN~phase} = \frac{\rm{time~to~photoionize~host~galaxy}}{\rm{fraction~of~optically~elusive~AGN}}
\end{equation}
In order to measure the typical duration of the AGN phase, we require estimates of the fraction of the AGN population lacking AGN lines (the optically elusive AGN or XBONGs) and the typical size of the narrow line region. We caution that this estimate is only true in a statistical sense as it is derived from population statistics, and that there are likely many systematics in both quantities -- we discuss those systematics later.

At a later time, once the black hole stops accreting, the ISM may continue producing AGN-like lines as the light echo from the past AGN travels out across the galaxy. At the largest scales, these light echoes have been seen as the quasar light echo "Hanny's Voorwerp" near IC 2497, and the Voorwerpjes (little Voorwerps) \citep{2009MNRAS.399..129L,2010A&A...517L...8R,2012AJ....144...66K, 2012MNRAS.420..878K} as well as in some quasars \citep{2013ApJ...763...60S, 2015MNRAS.449.1731D}. These Voorwerps constrain  the "shut-down" time scale to $10^{4}-10^{5}$ years via the geometry of their spatial extent and the recombination timescale of the narrow line emission. 

\begin{figure*}
\begin{center}

\includegraphics[angle=0,width=0.9\textwidth]{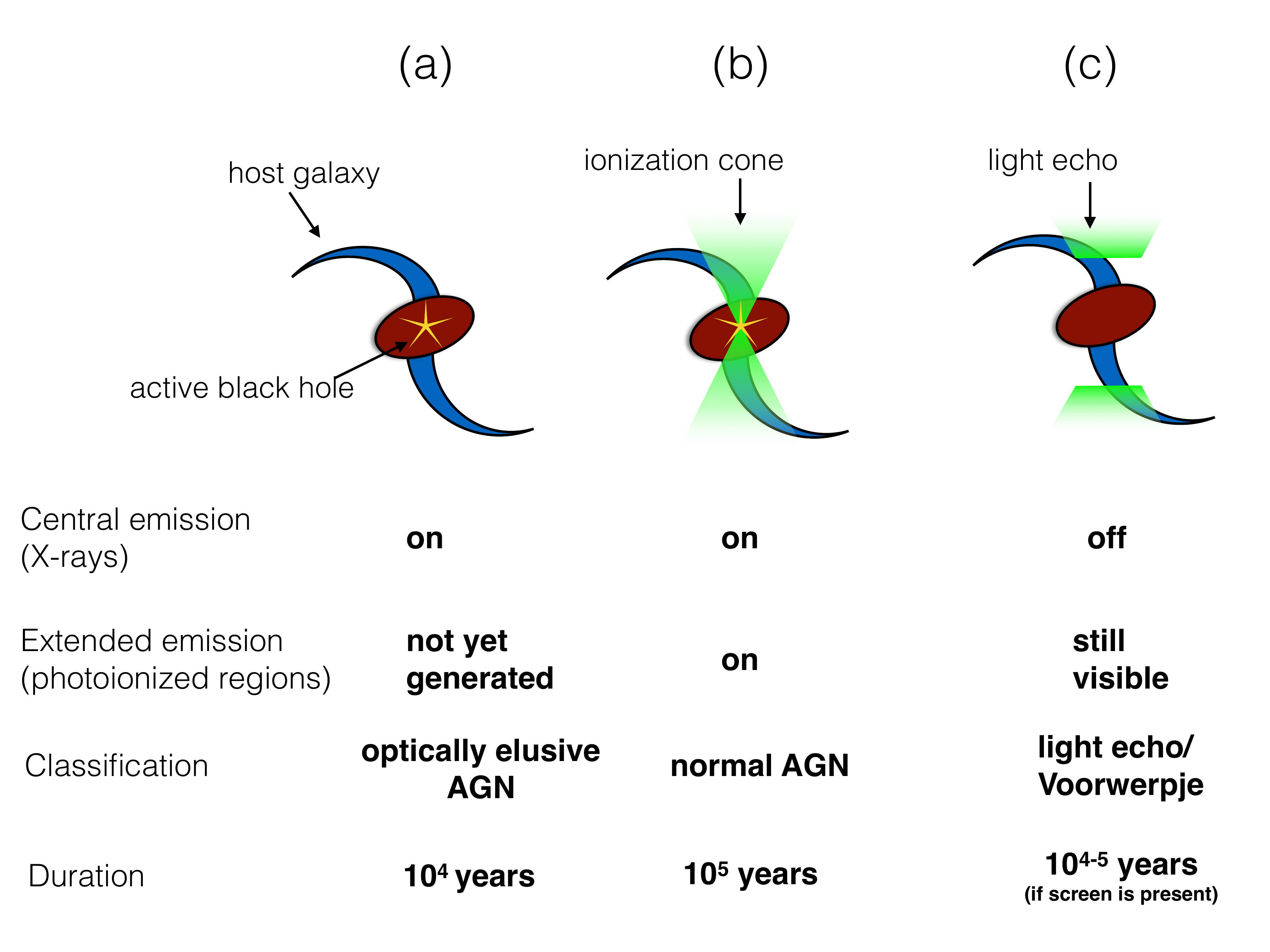}

\caption{A schematic of the life cycle of active galactic nuclei. Once the black hole starts accreting (a), the central engine begins emitting in X-rays, but the extended photoionized region on the scale of the galaxy has not yet been lit up; this is the XBONG phase. After the AGN central engine has photoionized the ISM of the galaxy, the AGN is visible as a ``normal" AGN with both nuclear X-ray emission and a galaxy-scale narrow line region. When the black hole stops accreting, the nucleus stops emitting X-rays, but the light echo from the AGN phase may still travel across the galaxy and to gas reservoirs beyond the galaxy: the galaxy may appear as a Voorwerpje. This cycle may repeat many times to make up a significant black hole growth episode.}

\label{fig:schematic}

\end{center}
\end{figure*}

We illustrate this AGN phase life cycle in a schematic form in Figure \ref{fig:schematic}.

\begin{table}
\begin{center}
\caption{Literature measurements of "optically elusive" and XBONG fractions}
\label{tab:samples}
\begin{tabular}{@{}lll}
\hline
 \multicolumn{1}{l}{AGN/galaxy} 	& \multicolumn{1}{l}{Optically elusive} 	& \multicolumn{1}{l}{Reference} 	\\
 \multicolumn{1}{l}{sample} 		& \multicolumn{1}{l}{fraction} 			& \multicolumn{1}{l}{} 	 			\\
\hline
\hline
 $BeppoSAX$					& 4\%								& \cite{2002ApJ...570..100L} \\	
 $Chandra$+SDSS				& 6\%								& \cite{2005AJ....129...86H} \\
 $Swift$-BAT						& 6\%								& \cite{2014ApJ...794..112S} \\
 $Swift$-BAT						& 5.6\%								& this work \\

 \hline
\end{tabular}
\\
\begin{flushleft}
\end{flushleft}
\end{center}
\end{table}


\begin{figure}
\begin{center}

\includegraphics[angle=0,width=0.49\textwidth]{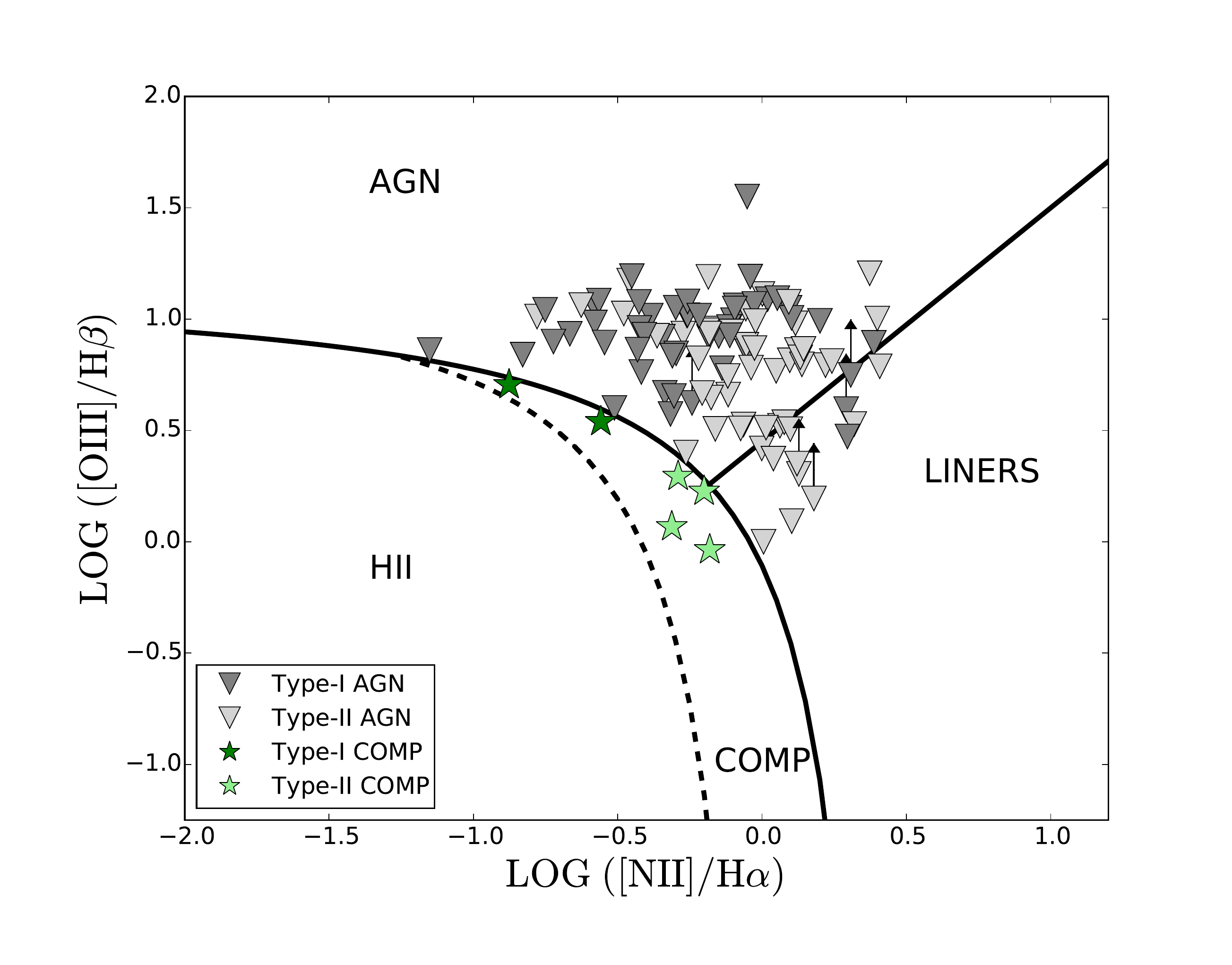}

\caption{Emission line ratio diagrams for a sample of local hard X-ray selected AGN observed by \textit{Swift} BAT and the Sloan Digital Sky Survey in the redshift range $0.0165<z<0.4$. 6 out of 107 AGN do not show a narrow line region yielding an XBONG fraction of 5.6\%. This fraction is comparable to those seen in other studies \citep{2002ApJ...570..100L, 2005AJ....129...86H, 2014ApJ...794..112S}. }

\label{fig:bpt}

\end{center}
\end{figure}

\subsection{Observations}

\subsubsection{Compilation of "optically elusive" AGN fractions from the literature}
If the life cycle for AGN phases outlined is correct, then the switch-on phase has already been seen: X-ray-selected AGN with host galaxies lacking obvious AGN lines. There is some diversity in terminology and definition of these objects in the literature. These AGN host galaxies can be described as "optically elusive" or "optically dull" AGN or as XBONGs (X-ray bright, Optically Normal Galaxies\footnote{The nomenclature for such objects is not uniform. Some studies reserve the term "XBONG" for X-ray selected AGN with host galaxies which show \textit{no} emission lines, as opposed to showing emission lines excited by processes other than AGN.}). These classes can sometimes contain AGN host galaxies which are red and dead and therefore do not have a sufficient cold ISM to photoionize and produce strong lines, but these objects make up a minority. Such AGN host galaxies have been reported repeatedly in the literature in a variety of surveys \citep{2002cosp...34E1560C, 2002ApJ...579L..71M, 2003MNRAS.344L..59M, 2007A&A...470..557C, 2005MNRAS.358..131G, 2006ApJ...645..115R, 2007A&A...466...31C, 2014ApJ...794..112S}. Independent of selection and definition, the fraction of such host galaxies is consistently around the $\sim5\%$ level as reported by various teams; we report these values in Table \ref{tab:samples}.

While the details of how AGN are selected on emission line diagrams vary, they share generic features which go back to the original \cite{1981PASP...93....5B} and \cite{1987ApJS...63..295V} which separate star-forming galaxies, AGN and other non-stellar ionization sources using line ratios. The most common diagram -- and the one we use in this paper -- is the standard \OIIIHb\ vs \NIIHa\ diagram. Star-forming galaxies are separated using an empirical line of \cite{2003MNRAS.346.1055K}. The "composite" region where star-formation and shocks can mix extend all the way to the extreme starburst line of \cite{2001ApJ...556..121K}. Seyfert AGN and LINERs (low ionization nuclear emission regions) beyond the extreme starburst line can further be separated \citep{2007MNRAS.382.1415S, 2006MNRAS.372..961K}. In general, AGN photoionized galaxies are identified by having high \OIIIHb\ line ratios, with a second line ratio such as \NIIHa\ separating out other processes such as shocks and evolved stellar populations.

For this paper, we add an independent measurement of the "optically elusive" fraction. We take the sample of hard X-ray selected AGN detected by \textit{Swift} Burst Alert Telescope (BAT) in the footprint of the Sloan Digital Sky Survey in the redshift interval $0.0165<z<0.4$ \citep{2009ApJS..182..543A, 2013ApJS..207...19B}. Using emission line diagnostics, we find that 6 out of 107 AGN, $5.6\%$ are classified as composite, meaning they are a mixture of AGN and star formation. We show the resulting classical \cite*{1981PASP...93....5B} \OIIIHb\ vs \NIIHa\ diagram in Figure \ref{fig:bpt}.

The SDSS fibers cover the nuclear region of the BAT AGN host galaxies so in our framework, the AGN in these galaxies just switched on and has begun to photoionize the nuclear region, but has not yet overwhelmed the star formation lines. Our measurement is consistent with that of \cite*{2014ApJ...794..112S} of a similar sample of \textit{Swift} BAT AGN, as well as with previous reports despite the heterogeneous nature of samples, selection techniques and definitions.

\subsubsection{Compilation of narrow line region sizes}
The switch-on time can be derived from the physical scale of the narrow line region (NLR), which to zeroth order scales as the half-light radius of the host galaxy. What do observations show that the NLRs of luminous AGN are large and at least on scales of several kpc \citep[e.g.][]{2003ApJ...597..768S}. Long-slit spectroscopy of obscured AGN by \cite{2014ApJ...787...65H} shows that the NLR reaches the physical scale of the entire host galaxy as the AGN luminosity increases and tops out, leading to a flattening of the relationship between NLR size and AGN luminosity. Integral field unit studies of AGN host galaxies similarly show large scale NLRs \citep{2015MNRAS.446.2186M}. Based on these observations of NLR sizes, we can take the half-light radius of the host galaxy of a luminous AGN as a reasonable proxy for the expected NLR size. 

For this paper, we take the typical half-light radius of the \textit{Swift} BAT AGN host galaxy sample of $\sim 5$ kpc, which corresponds to $\sim 16,000$ light years. To be conservative, we assume that the typical radius of the NLR is somewhat smaller and take a switch on time of $\sim 10,000$ years.

\subsection{Estimate of the AGN phase lifetime}
We now have estimates of the two quantities that go into the calculation of the AGN phase lifetime. We take the $5\%$ optically elusive fraction, and the 10,000 year switch on time scale: this yields an AGN phase lifetime of:
\begin{equation}
t_{\rm AGN~phase} \sim \frac{10,000}{0.05} \sim 200,000~\rm{years}
\end{equation}
We stress that this estimate should be seen as an order of magnitude estimate, and that it only applies in a statistical sense. Nevertheless, we believe this typical lifetime has significant consequences for our understanding of AGN variability and the role AGN play in galaxy evolution. Before we address these, we discuss some of the major caveats to this estimate.

\subsection{Caveats}
\subsubsection{Is there sufficient gas in the host galaxies?}
\label{sec:gaspoor}
In order to light up a narrow line region, the AGN radiation needs a sufficiently dense screen of cold gas in the host galaxy to illuminate. One of the hypotheses for the nature of XBONGs is that they are gas-poor, red sequence galaxies. Correspondingly, their spectra show no strong emission lines \textit{of any kind} while still featuring AGN X-ray emission. For the more distant AGN/XBONG samples, this may be a concern, but for the \textit{Swift} BAT sample, we can rule out this explanation for most host galaxies: as \cite{2011ApJ...739...57K} report, very few \textit{Swift} BAT host galaxies are red and gas poor. If such gas-poor AGN hosts\footnote{The XBONGs, in some definitions} are a significant fraction of the total "optical elusive" AGN fraction, this would increase the inferred AGN lifetime. A 50\% contamination would double the inferred AGN lifetime. In our sample of \textit{Swift} BAT AGN, only one object features no emission lines, which implies that this is not a major bias.

\subsubsection{Are XBONGs/optically elusive AGN not simply dusty?}
\label{sec:dusty}
Another explanation for the apparent lack of AGN lines in the optical spectra is that XBONGs/optically elusive AGN are particularly dusty and so their AGN lines are not apparent in the optical spectrum. Searches for AGN lines using near-infrared spectroscopy of \textit{Swift} BAT AGN host galaxies by \cite*{2014ApJ...794..112S} find that most XBONGs lack AGN-like lines. Even in the case of two objects with near-IR lines consistent with AGN, star-formation remains a possibility. If a significant fraction of optically elusive AGN are in fact normal AGN with dust obscured NLRs, this would increase the inferred AGN lifetime. As with Section \ref{sec:gaspoor}, a 50\% contamination would double the inferred lifetime.

\subsubsection{Is the assumption of "light travel time" too simplistic?}
We assume that we can measure the switch on time using the light travel time across the host galaxy as the relevant time scale. This is naturally a simplification as the switch on process is gradual. More importantly, we are neglecting the well-studies physics of photoionization which certainly modulates the mapping from light travel time to appearance of narrow lines.

How much of a delay can such effects cause? In the central parts of a galaxy, the electron density is high enough that the recombination timescale is very short. For typical values of density and temperature, the \Hb\ recombination timescale is about $\sim 200$ years (see, e.g., \citealt{2013ApJ...779..109P}). It is only in the outer regions of a galaxy where the density drops significantly that the recombination timescale becomes substantial (as is the case with some Voorwerps, see \citealt{2012AJ....144...66K, 2012MNRAS.420..878K}).  Other timescales which contribute to the delay, such as the time needed to significantly increase the ionization level, or to excite a significant fraction of ions, are expected to be shorter than the recombination timescale. The host galaxy ISM is likely already largely photoionized due to star formation, so all the AGN ionization front needs to do is change the relative populations in the ions present. In any case, any additional delay between the AGN photoionization front moving though the host galaxy and exhibiting AGN lines would increase the "optically elusive" AGN fraction artificially, and therefore further \emph{decrease} the inferred lifetime.

\subsubsection{Could aperture effects be significant?}
In a host galaxy with significant star formation, the emission lines from star formation could overwhelm the AGN lines, moving the spectroscopic classification out of the AGN region and into the composite region. This is a particular concern for SDSS fiber spectra as the large physical footprint of the fiber can include such a signal from star formation. \cite{2010ApJ...711..284S} showed using simulations that this dilution effect is significant for low-luminosity AGN around \LOIII\ $\sim10^{39}$ \ergs\, but ceases to be important for brighter AGN around \LOIII\ $\gtrsim 10^{40.5}$ \ergs. Since the \textit{Swift} BAT AGN are luminous, this dilution effect is unimportant.

Furthermore, not all optically elusive samples are based on fiber spectra: both the \cite{2002ApJ...570..100L} and the \cite{2014ApJ...794..112S} studies use long-slit spectroscopy and find similar optically elusive fractions as those based on fiber spectra. Ultimately, integral field unit observations will resolve this. AGN during the switch-on phase should exhibit AGN-like lines in the nucleus and lack them at larger radii despite hosting an AGN sufficiently luminous (based on X-ray observations) to photoionize the entire galaxy.

As with the caveats in Section \ref{sec:gaspoor} and \ref{sec:dusty}, a significant fraction of AGN misclassified due to aperture effects would increase the inferred AGN lifetime. We note that while all the caveats discussed here could affect the inferred AGN lifetime, the magnitude of the implied changes are on the order of a factor of $\sim2$ and thus have only a moderate effect on our order-of-magnitude estimate of the AGN lifetime.

\begin{figure*}
\begin{center}

\includegraphics[angle=0,width=0.9\textwidth]{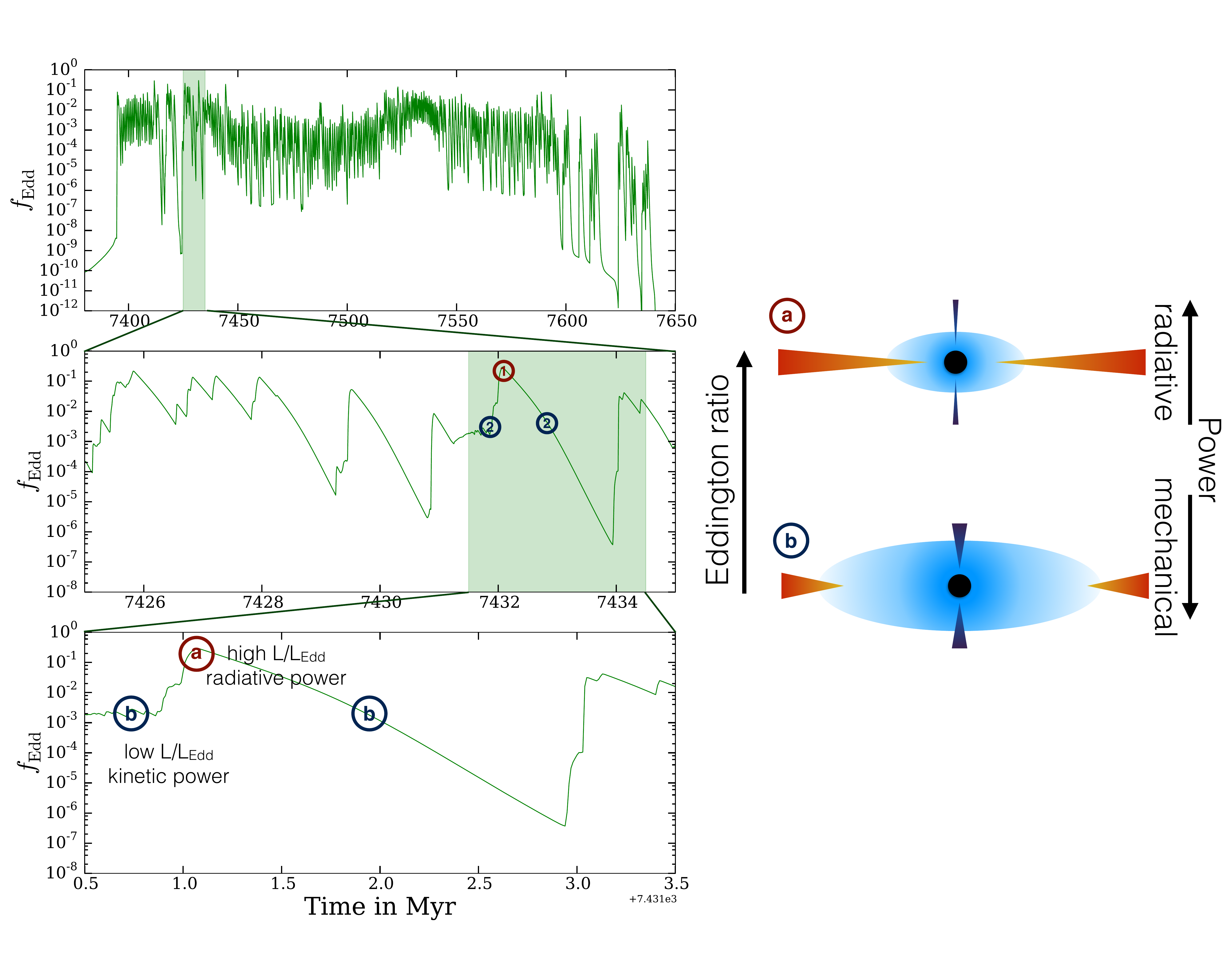}

\caption{On the \textit{left}, we show a simulated AGN light curve from \citet{2011ApJ...737...26N}. The \textit{top} panel covers a 250 Myr AGN phase sub-divided into many shorter $\sim10^{5}$ year bursts. The \textit{middle} and \textit{bottom} panel show zoom-ins of this light curve. This simulated light curve illustrates how the growth history of this black hole is made up of numerous short near-Eddington bursts lasting $\sim10^{5}$ years. On the \textit{right}, we show a schematic view of an accreting massive black hole following the illustration of \citet*{2007A&ARv..15....1D} for X-ray binaries and adapted for AGN by \citet{2012NewAR..56...93A}. During the high-Eddington peaks (\textit{a}), the black hole shines as a Seyfert/quasar with an optically thick, geometrically thin \citet{1973A&A....24..337S} disk, while during low-Eddington troughs (\textit{b}), the black hole transitions to a more geometrically thick, optically thin accretion flow \citep[e.g.][]{1994ApJ...428L..13N}.}

\label{fig:novak}

\end{center}
\end{figure*}

\section{Discussion}
\label{sec:discussion}

We now briefly discuss several consequences of a characteristic lifetime on the order of $10^{5}$ years for AGN phases. We note that our argument and the resulting AGN lifetime is in broad agreement with other studies, as discussed below:

\cite{1997ApJ...482L...9S} presented a simple model for the quasar luminosity function based on the assumption that quasar activity is driven by thermal-viscous instability in the accretion disk. This model yields short bursts of comparable duration to our measurement driven by such instabilities and the resulting model luminosity function matches the observed data very well.

\cite{2014ApJ...782....9H} have argued that the shorter characteristic time scale of AGN relative to star formation in galaxies can account for the \textit{apparent} disconnect between star formation and black hole accretion. They propose a simple model for the distribution of Eddington ratios with a functional form of a power law with an exponential cutoff at high Eddington ratios and allow AGN to vary following this distribution. If star-forming galaxies (\textit{i.e.} those galaxies with gas) follow this prescription for black hole accretion, the \cite{2014ApJ...782....9H} model produces a reasonable AGN luminosity function and matches the observed relationship between star formation rate and black hole accretion rate. The model does not constrain the actual AGN phase lifetime, \textit{i.e.} the time spent at high Eddington ratio in any individual burst; our observational argument now constrains this time.

High resolution simulations have similarly indicated short AGN lifetimes: \cite{2011ApJ...737...26N} perform very high resolution hydrodynamical simulations of feeding and feedback in elliptical galaxies, resulting in short $10^{5}$ year AGN bursts. Similarly, \cite{2011ApJ...741L..33B} and \cite{2013MNRAS.434..606G} find  short high-Eddington AGN bursts in simulations of high redshift gas-rich disks with black holes. In these simulation cases, the enormous short time scale variability is driven by clumpy and highly variable inflow of gas in the very center of the galaxy.

All these separate approaches are converging on a picture where short bursts are the norm for black hole growth. It remains however unclear if the burst scale is driven by the fueling mechanism (such as the clumpiness of the material driven towards the center), or by instabilities in the accretion disk. We cannot resolve this issue here, but the fact that we can observationally support short individual AGN phases has a number of interesting consequences:

\subsection{Black holes accrete mass through many short phases}
The first major consequence is that each individual black hole growth phase contributes only very little to the total mass growth of black holes. The \cite{1982MNRAS.200..115S} argument implies characteristic growth times of black holes on the order of 100 Myr to 1 Gyr \citep[see also][]{2002MNRAS.335..965Y} for reasonable radiative efficiencies $\eta$. In order to reconcile the short lifetime of $10^{5}$ years with a total growth time of $10^{8-9}$, the total growth of massive black holes must consist of \textit{many} (~100-1000) such short accretion bursts.

\subsection{Connection to shorter- and loger-term variability}
The characteristic time scale of $10^{5}$ years for the AGN phase must be placed in the context of both shorter- and longer-term variability as we know AGN exhibit both. On the shorter term side, there is a rich variety of observations on the hour to years time scale as many local AGN are monitored and sometimes exhibit significant changes in luminosity. This shorter-term would be superimposed on the total duration of a $10^{5}$ year phase, as can be seen e.g. in the simulated light curve of \cite{2011ApJ...737...26N} show in Figure \ref{fig:novak}. Taking several $10^{5}$ year phases, it seems unlikely that these would be randomly distributed. Rather, it is likely that a large number of 100 to perhaps 10,000 make up a longer growth phase of $10^{7}$-$10^{9}$ years. To put it another way, what we previously considered to be a $\sim10^{8}$ year AGN phase, perhaps triggered by a major merger, should be divided into $\sim1000$ short bursts. In this view, AGN "flicker" in the same sense as a fluorescent lamp\footnote{This very useful analogy was first proposed by R. Hickox.} does.

\subsection{Rapid flickering in AGN may imply changes in accretion state, similar to X-ray binaries}
The flickering behavior for AGN connects the behavior of massive black holes to that of stellar mass black holes. X-ray binaries exhibit often rapid changes in Eddington ratio and accretion state which changes their appearance and their impact on their environment. The idea that the accretion physics in X-ray binaries and AGN is similar, except that AGN black holes are more massive and therefore change on significantly longer time scale has been suggested a number of times \citep[e.g.][]{2003MNRAS.345L..19M, 2004A&A...414..895F, 2006MNRAS.372.1366K,2006Natur.444..730M,2010ApJ...724L..30S}. In this framework, the different classes of AGN are closely related: black holes at a given mass rise and fall in Eddington ratio, leading to appearance either as a quasar or a radio galaxy, respectively. 

\subsection{Time scale implications for AGN feedback}
Dividing a single sustained AGN phase into numerous short phases alternating between high-Eddington bursts and lower-Eddington troughs points to a new way in which the black hole-galaxy connection works. During each "cycle", the mode via which energy is injected into the system changes: during the high-Eddington peak, the AGN will produce predominantly radiative energy, while during the low-Eddington troughs, the change in accretion mode means that most of the energy is injected in kinetic form \citep{2007A&ARv..15....1D, 2012NewAR..56...93A}. The AGN flickering cycle thus alternates between radiative and kinetic modes on a short time scale compared to the dynamical time of the host galaxy. This "jackhammer" effect could be important in driving AGN outflows and feedback. Even if each individual cycle only injects a small amount of energy, repeating the cycle 100-10,000 times could add up to sufficient amounts to unbind the gas reservoir of the host galaxy as required for AGN feedback. In order to approach the binding energy of a massive galaxy of $\sim10^{60-61}$ erg, each cycle would only have to inject $\sim10^{56-58}$ ergs.

The short AGN life cycle makes linking AGN activity to outflow phenomena much more challenging. The low-Eddington phase may be difficult or impossible to detect in star-forming host galaxies and thus the galaxy may not be classified as an AGN, despite the AGN being active. Similarly, a kinetic outflow driven by the low-Eddington phase may persist, perhaps coasting ballistically, until the next high-Eddington burst. Searches for AGN outflows generally target bright AGN, and so might erroneously associate observed outflows with the bright phase, when in fact, the previous faint, low-Eddington phase was responsible.

A short lifetime (compared to the total growth time) for the AGN phase causes a time scale problem relative to the galaxy. The AGN central engine operate on timescales comparable only to the central dynamical time of the host galaxy, and orders of magnitude faster than the dynamical time of the whole host galaxy ($t_{\rm dyn} \sim 10^{8}$ years for a massive galaxy). The AGN can switch on and off many times on the time the host galaxy at large is able to react to. This is exacerbated by the fact that most tracers of the host galaxy star formation rate and stellar age are dependent on tracers whose time resolution is much worse than the $10^{5}$ time scale of the AGN. H$\alpha$ is driven by photoionization of OB stars and from the lifetime of these stars alone yields a time "smoothing kernel" of at least $10^{6}$ years. Ultraviolet emission from young stars and reprocessed infrared extend even further in time as they respond to young stellar populations. Spectral features such as Balmer absorption lines or the 4000\AA\ break have even worse time resolution. This mismatch in time scales may explain why it is so difficult to link AGN to galaxy evolution, and suppression of star formation in particular. The AGN lives and acts of a much shorter time scale than can be reconstructed from stellar populations.

\subsection{The Milky Way black hole}
The recent activity of the black hole in the Milky Way center can be revisited in light of a short AGN flickering cycle. X-ray light echoes in the Galactic Center show moderate-luminosity flares in the last few centuries \citep{2004A&A...425L..49R, 2010ApJ...714..732P}, though these were likely too short and not sufficiently luminous to generate narrow lines across the Milky Way. \cite{2013ApJ...778...58B} show that the Magellanic Stream near the Milky Way was photoionized by an AGN event in the Galactic center some 1-3 Myr ago, and the duration of the burst is estimated to be 100-500 kyr - the same order of magnitude as the lifetime derived in this paper. \cite{2013ApJ...778...58B} propose this Galactic Center outburst as the cause of the \textit{Fermi} bubble \citep{2010ApJ...724.1044S}. If this is the case, the flickering cycle exhibited by other AGN should leave similar remnants as the \textit{Fermi} bubble around other galaxies as their black holes switch on and off.

\section{Summary}
We have presented an observational constraint for the typical AGN phase lifetime. The argument is based on the time lag between an AGN central engine switching on and becoming visible in X-rays, and the time the AGN then requires to photoionize a large fraction of the host galaxy. Based on the typical light travel time across massive galaxies, and the observed fraction of X-ray selected AGN without AGN-photoionized narrow lines, we estimate that the AGN phase typically lasts $\sim10^{5}$ years. This short lifetime implies that black holes grow via many such short bursts and that AGN therefore "flicker" on and off. We discuss some consequences of this flickering behavior for AGN feedback and the analogy of X-ray binaries to AGN.


\section*{Acknowledgements}
We thank the anonymous referee for helpful comments. We are grateful to G. Novak for sharing his simulated light curves. We thank Benny Trakhtenbrot, Ezequiel Treister, Meg Urry and Bill Keel for useful discussions. KS, MK and LS gratefully acknowledge support from Swiss National Science Foundation Professorship grant PP00P2\_138979/1 and MK support from SNSF Ambizione grant PZ00P2\_154799/1. This research has made use of data from the Sloan Digital Sky Survey and the NASA Swift satellite. This research has made use of NASA's ADS Service.

\bibliographystyle{mn2e}

\bsp

\label{lastpage}

\end{document}